\documentclass[a4paper]{jpconf}
\usepackage{color}
\usepackage{graphicx}
\begin{document}
\title{Jet measurements in polarized p+p collisions at STAR at RHIC}

\author{Xuan Li$^{[1]}$ for the STAR Collaboration}

\address{1) Physics Department, Temple University, Philadelphia, 19122, USA}

\ead{xuanli@rcf.rhic.bnl.gov}

\begin{abstract}
Jet production in polarized $p+p$ collisions at $\sqrt{s} = 200$ GeV and $\sqrt{s} = 500$ GeV provides a powerful probe to study gluons inside the proton. The Solenoidal Tracker at RHIC (STAR) has the capability, with nearly full azimuthal ($2\pi$) coverage, to reconstruct jets at mid-rapidity ($|\eta| < 1$). The latest STAR inclusive jet longitudinal double-spin asymmetry $A_{LL}$ measured in 200 GeV $p+p$ collisions provides better constraints on the polarized gluon distribution $\Delta g(x)$ for $0.05<x<0.2$ than previous measurements. A recent global QCD fit (DSSV 2014) which includes the 2009 RHIC results provides the first evidence of non-zero gluon contribution to the proton spin. A new inclusive jet cross section using the anti-$k_{T}$ algorithm provides potential insights into the unpolarized gluon distribution function, and the new inclusive jet $A_{LL}$ measurement in 510 GeV $p+p$ collisions shows consistent $x_{T}$ scaling with the 200 GeV result. Future measurements with continuing high energy polarized proton-proton running at $\sqrt{s} = 500$ GeV at RHIC and detector upgrades in the forward direction will explore the gluonic contribution to the proton spin at low $x$ region.
\end{abstract}

\section{Introduction}
One of the main objectives for the Relativistic Heavy Ion Collider (RHIC) spin program is to quantitatively determine the polarized gluon distribution $\Delta g(x, Q^{2})$ through longitudinally polarized proton-proton collisions. Since 2009, RHIC has collided longitudinally polarized proton beams at $\sqrt{s} = 200$ GeV and $\sqrt{s} = 500/510$ GeV. The Solenoidal Tracker at RHIC (STAR) has both tracking detectors and electromagnetic calorimeters with nearly full azimuthal coverage, which allows for final states, such as jets, to be well reconstructed at mid-rapidity ($|\eta|<1$). In leading order QCD calculations, mid-rapidity jet production is dominated by quark+gluon and gluon+gluon interactions. Therefore hard scattered gluons can be probed by jet measurements at STAR in $p+p$ collisions. The STAR collaboration has submitted for publication the measurement of the inclusive jet double spin asymmetry $A_{LL}$ in 200 GeV $p+p$ collisions based on data collected in the 2009 running period \cite{2009_jet_All}. A global QCD fit, which includes the 2009 STAR jet result, extracts a non-zero gluon spin contribution ($\Delta g(x,Q^{2})$) to the proton in the $0.05<x<0.2$ region \cite{dssv2014}. This contribution reports a new inclusive jet cross section measurement at STAR from the 2009 data set. The new measured inclusive jet cross section in data is consistent with the next-leading order (NLO) perturbative QCD calculations within uncertainties for the $0.1<x<0.5$ region. 

The uncertainty of $\Delta g(x, Q^{2})$ remains large for the low $x$ region ($x<0.05$). STAR can begin to probe some of this lower $x$ region by measuring asymmetries of final states at different center of mass energies, such as the inclusive jet $A_{LL}$ in 510 GeV $p+p$ collisions, and by moving to more forward rapidities.

\begin{figure}[t]
\centerline{\includegraphics[width=28pc]{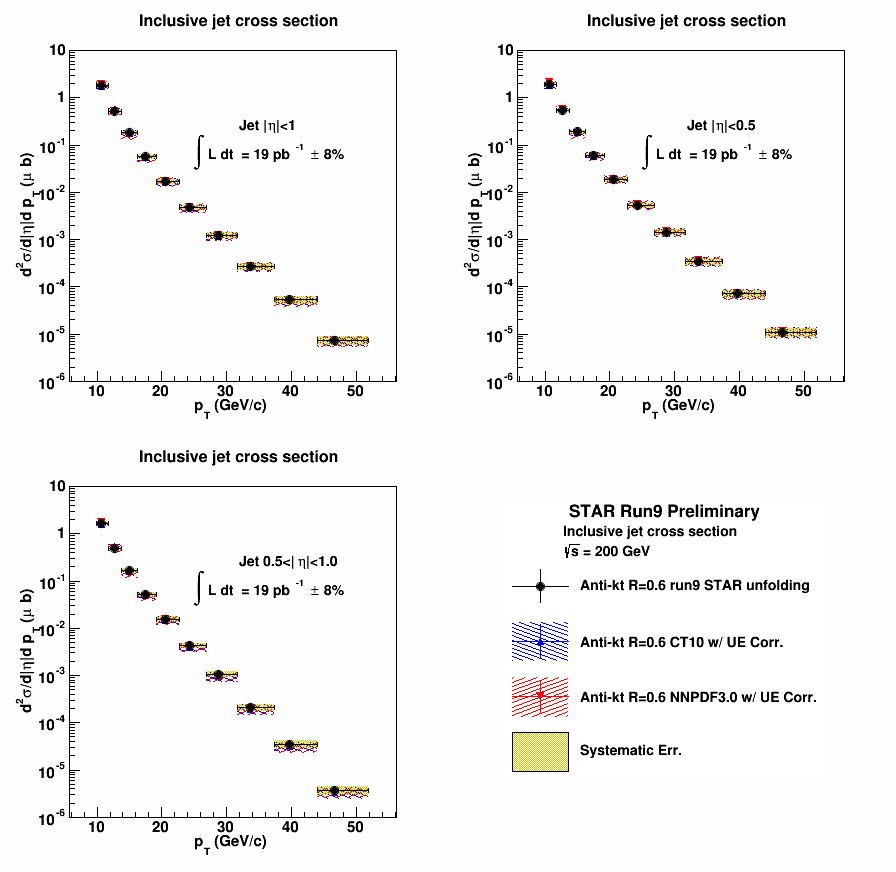}}
\caption{ \it 2009 inclusive jet cross section preliminary result in 200 GeV $p+p$ collisions within jet $|\eta|<1.0$ (top left panel), $|\eta|<0.5$ (top right panel) and $0.5<|\eta|<1.0$ (bottom left panel) pseudorapidity regions. The results are compared with underlying event and hadronization corrected NLO pQCD calculations with CT10 (blue) and NNPDF3.0 (red) PDF sets with factorization and renormalization scaling uncertainties.}
\label{2009jetXsec}
\end{figure}

\section{Inclusive jet cross section in 200 GeV p+p collisions}
A primary goal of STAR in the 2009 $p+p$ run at $\sqrt{s} =200$ GeV is to study the dynamics and the spin structure of the proton via jet measurements. STAR uses Jet Patches (JP) \cite{starjet_2012}, which are defined as groups of adjacent calorimeter towers inside the electromagnetic calorimeter used to trigger on jet events. The anti-$k_{T}$ jet algorithm \cite{anti_kT} with a cone radius of R=0.6 is applied in the 2009 inclusive jet analysis, to replace the mid-point cone algorithm used in earlier analysis \cite{2003jet}. The anti-$k_{T}$ algorithm reduces the underlying event contribution and pile up backgrounds to the reconstructed jets. Particle level jets which are based on final state particles and parton level jets, which are formed from hard scattered partons (excluding those from the underlying events and beam remnants), can be reconstructed in the simulation. Jets reconstructed from the detector response (e.g. charged tracks or neutral energy depositions) in data are defined as detector level jets. Detector level jets can be reconstructed in simulations as well with the same trigger and detector conditions as in the data. In simulation, a detector level jet and a parton or particle level jet are considered to match each other when the distance between the two in $\eta-\varphi$ space is minimal and less than the jet cone radius. 

\begin{figure}[t]
\centerline{\includegraphics[width=28pc]{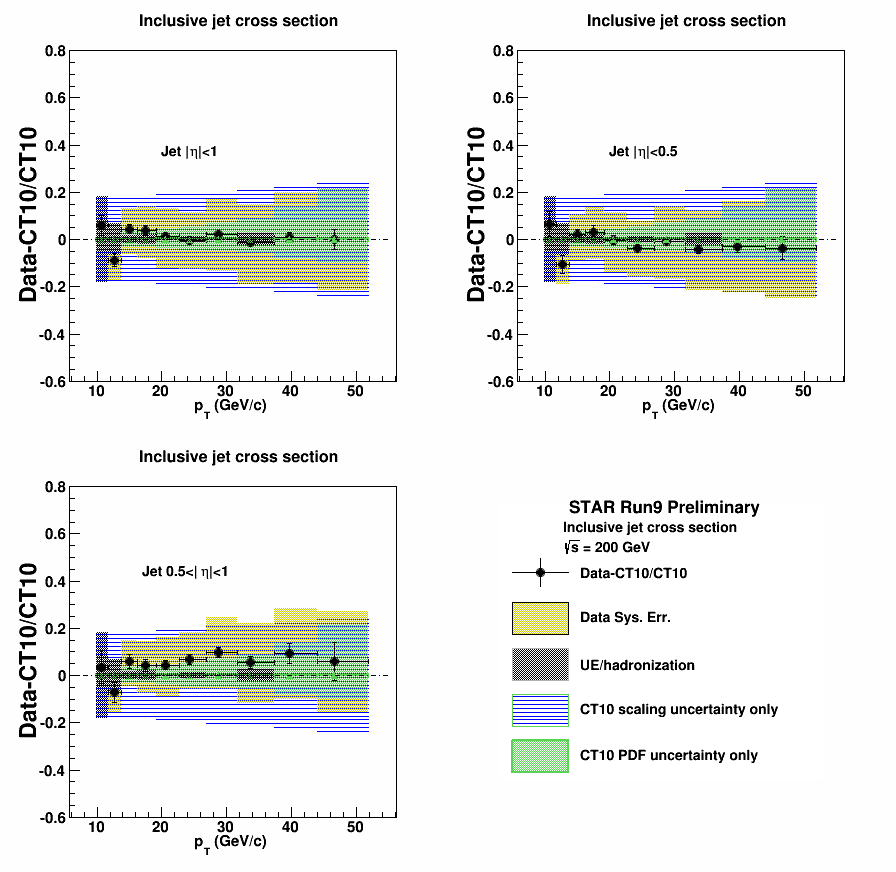}}
\caption{ \it Preliminary results of the measured inclusive jet cross section compared with NLO pQCD theory calculations using CT10 PDFs for jet $|\eta|<1.0$ (top left panel), $|\eta|<0.5$ (top right panel) and $0.5<|\eta|<1.0$ (bottom left panel) in 2009 200 GeV $p+p$ collisions. The yellow band stands for the systematic error, the grey band stands for the underlying event and hadronization correction on the theory calculations, the blue band stands for the CT10 scaling uncertainty and the green band stands for the CT10 PDF uncertainty.}
\label{dataCT10}
\end{figure}

For the inclusive jet cross section study, the detector jet transverse momentum $p_{T}$ measured in the data is corrected to the particle jet $p_{T}$ via unfolding. The primary sources of the systematic errors include the tracking resolution, the tracking inefficiency and the electromagnetic calorimeter resolution. The dominant systematic uncertainty of this measurement comes from the calibration of the electromagnetic calorimeter. Figure \ref{2009jetXsec} shows the preliminary result of the 2009 inclusive jet cross section versus particle jet $p_{T}$ for jets measured in $|\eta|<1.0$ (top left panel), $|\eta|<0.5$ (top right panel) and $0.5<|\eta|<1.0$ (bottom left panel) pseudorapidity regions in 200 GeV $p+p$ collisions. The total luminosity determined by the Beam Beam Counter (BBC) is 19 $\textrm{pb}^{-1}$ with $8\%$ scaling uncertainty. Results in data are compared with NLO pQCD calculations with the CT10 \cite{CT10} and the NNPDF3.0 \cite{NNPDF} PDF sets after underlying event and hadronization corrections are applied to the pure NLO calculations. The latter is determined by PYTHIA 6.426 simulations by studying the difference between the particle level jet cross section and the hard scattering parton jet cross section for each jet $p_{T}$ bin. The maximum value of the hadronization and underlying event correction is around $5\%$. In addition to this, a ratio is defined as the difference between the measured and theory calculated inclusive jet cross section normalized by the theory calculations. Figure \ref{dataCT10} shows this ratio using the CT10 PDF set for the theory calculations. Data are in good agreement with the NLO pQCD calculations when using the CT10 PDF set. The gluon distribution function $g(x,Q^{2})$ has large uncertainties in the high x region ($x>0.1$). The new STAR inclusive jet cross section measurements are divided into three different pseudo-rapidity regions ($|\eta|<1.0$, $0.5<|\eta|<1.0$ and $|\eta|<0.5$) to access the unpolarized gluon distribution functions in different x regions. This result is expected to provide new input to constrain $g(x,Q^{2})$ in the RHIC sensitive x region ($0.1<x<0.5$). 

\section{Inclusive jet longitudinal double spin asymmetries $A_{LL}$ in 200/510 GeV p+p collisions}
\begin{figure}[t]
\begin{minipage}{16pc}
\includegraphics[width=14.5pc]{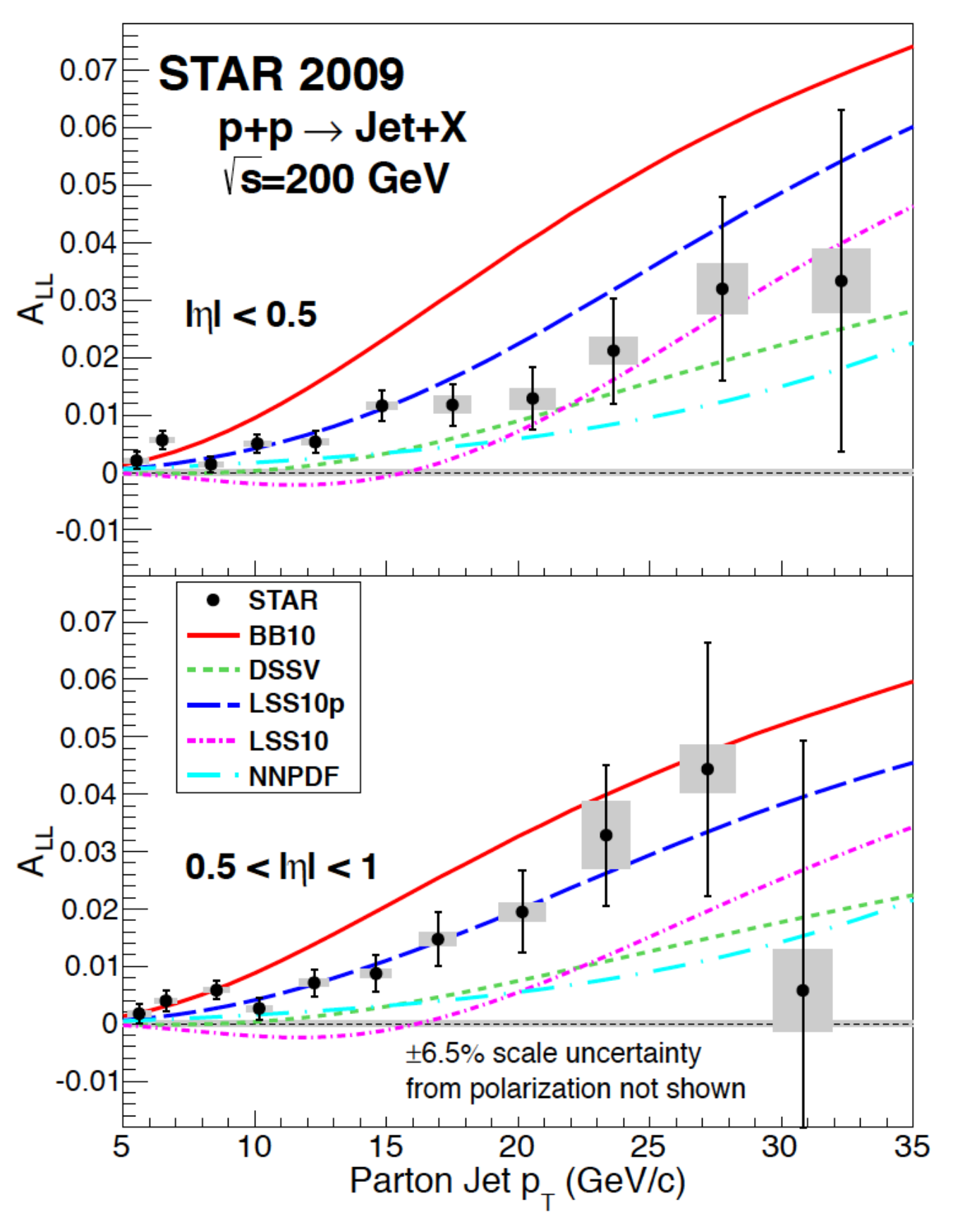} %\hspace{0.5pc}
\caption{\label{2009jet_plot} \it STAR 2009 inclusive jet $A_{LL}$ versus parton jet $p_{T}$ in 200 \textup{GeV} $p+p$ collisions \cite{2009_jet_All}. The $A_{LL}$ of the inclusive jets with $|\eta|<0.5$ is shown in the top panel, and the results with $0.5<|\eta|<1.0$ are shown in the bottom panel. The error bars are statistical, while the gray boxes represent the systematic uncertainties.}
\end{minipage} 
\hfill\begin{minipage}{17pc}
\includegraphics[width=14pc]{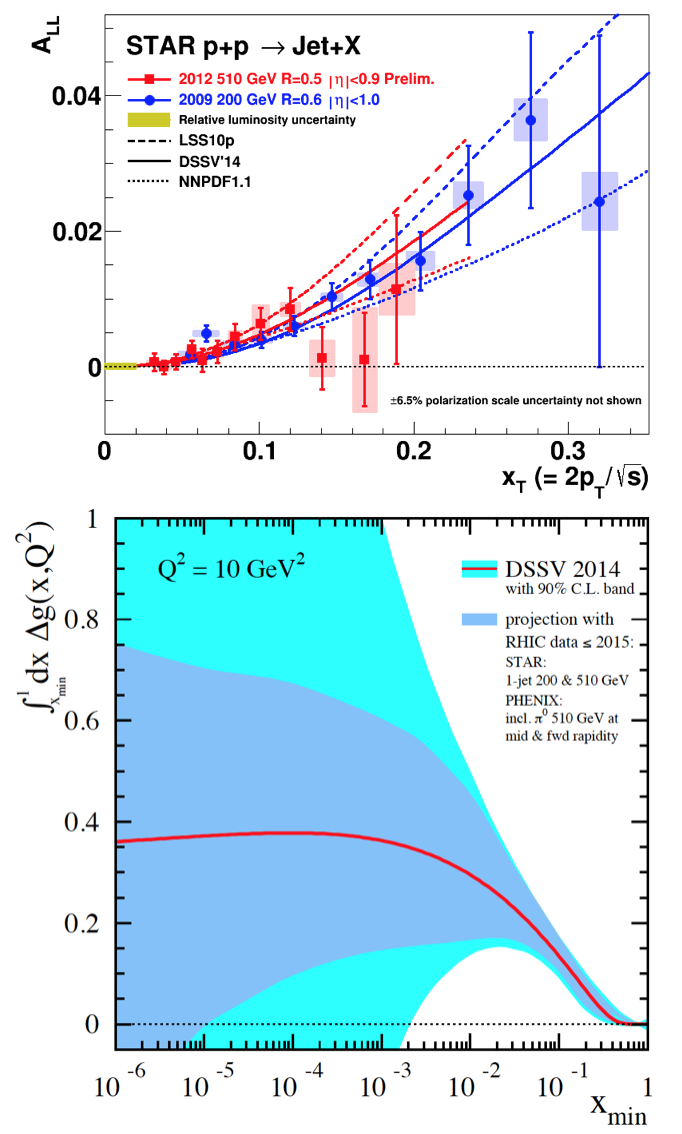}
\caption{\label{2012jet_plot} \it Top panel: STAR 2009/2012 inclusive jet $A_{LL}$ versus $x_{T} = 2p_{T}/\sqrt{s}$ in 200/510 \textup{GeV} $p+p$ collisions \cite{rhic_2015}. 2009 inclusive jet $A_{LL}$ with $|\eta|<1.0$ is shown in blue and the 2012 result with $|\eta|<0.9$ is shown in red. Bottom panel: Recent global analysis from the DSSV group on the running integral of $\Delta g$ as a function of $x_{min}$ at $Q^{2} = 10$ \textup{GeV}$^{2}$ \cite{rhic_2015}. Inner and outer bands stands for the 90 $\%$ C.L. uncertainties with and without including the projection of up to 2015 RHIC data.}
\end{minipage}
\end{figure}

The double spin asymmetry $A_{LL}$ of jets probes the helicity distribution of involved partons. Therefore, the gluon polarization contribution to the proton can be directly probed by measuring inclusive jets or di-jets. From the 2009 RHIC run, STAR collected 200 GeV $p+p$ data which contain 20-fold more statistics than the previous result \cite{2006_jet_ALL}. Figure \ref{2009jet_plot} shows the 2009 STAR mid-rapidity inclusive jet double spin asymmetry $A_{LL}$ versus the parton jet $p_{T}$ in 200 GeV proton-proton collisions \cite{2009_jet_All}. The measured detector jet $A_{LL}$ values are corrected to the parton jet $A_{LL}$ in each jet $p_{T}$ bin according to calculations done in the simulation with polarized/unpolarized PDFs. The systematic error on the product of the polarization of the two beams is $6.5\%$. The jet acceptance is divided into $|\eta|<0.5$ (top panel) and $0.5<|\eta|<1$ (bottom panel) to access different $x$ regions. The colored lines represent the NLO pQCD inclusive jet $A_{LL}$ calculations with different polarized parton distributions used as inputs \cite{dis,dis1,theory1,theory2,theory5}. The data points fall between the LSS10p \cite{dis} and the DSSV \cite{theory1} model predictions.

The DSSV group is the first to apply a global QCD fit on the combined data samples from DIS, Semi-Inclusive DIS (SIDIS) and RHIC $p+p$ experiments. Recently, the NNPDF group performed a global QCD fit to extract polarized PDFs based on the polarized DIS and polarized RHIC $p+p$ data. The extracted gluon polarization contribution from the latest NNPDF fit in the range $0.05 < x < 0.2$ is $0.17 \pm 0.06$ including the 2009 STAR jet data \cite{dis1}. The latest DSSV update, which includes the 2009 STAR inclusive jet $A_{LL}$ \cite{2009_jet_All} and 2009 PHENIX inclusive $\pi^{0}$ $A_{LL}$ \cite{Phenix_pi0_ALL}, exhibits a positive polarized gluon helicity function \cite{dssv2014}. The fitted integral of the gluon helicity function in this $x$ region ($\Delta G = \int_{0.05}^{1} \! \Delta g \, \mathrm{d}x (Q^{2} = 10$ GeV$^{2}) = 0.2^{+0.06}_{-0.07}$ within $90 \%$ confidence level) indicates a first non-zero gluon polarization contribution to proton observed at RHIC.

Higher energy (i.e. 500/510 GeV) $p+p$ collisions at RHIC provide a means to access lower x gluons. In 2012, STAR recorded around 50 $\textrm{pb}^{-1}$ luminosity of longitudinal polarized $p+p$ data with an average beam polarization of $53\%$. The jets were reconstructed with the anti-$k_{T}$ jet algorithm. The cone radius R was reduced to 0.5 in order to help suppress the backgrounds from pile up and enhance the detector-level jet and the parton-level jet association probabilities. The top panel of figure \ref{2012jet_plot} shows the STAR preliminary result of the inclusive jet longitudinal double spin asymmetries $A_{LL}$ measured in 2012 510 GeV $p+p$ collisions in comparison with the 2009 inclusive jet $A_{LL}$ as a function of $x_{T}$. The 2012 510 GeV inclusive jet $A_{LL}$ result is in agreement with the 2009 200 GeV result in the overlapping region, and extends to the lower $x_{T}$ region.

RHIC operated another 200 GeV longitudinally polarized proton-proton run in 2015. The inclusive jet $A_{LL}$ measurement from the combined statistics from 2009 to 2015 will provide better constraints with improved statistics on the gluon polarization in the same $x$ region as the 2009 result. In addition to the 200 GeV longitudinal proton-proton run, RHIC also performed 510 GeV longitudinal proton-proton collisions with around $600 \ \textrm{pb}^{-1}$ delivered luminosity in 2012 and 2013. The di-jet production can access low x gluons via asymmetric partonic scattering. Gluon polarization in the lower $x$ region will be probed by the ongoing 2012/2013 di-jet analysis at the higher center of mass collision energy. Impacts from these data sets together with the PHENIX forward $\pi^{0}$ $A_{LL}$ projections are shown in the bottom panel of Figure \ref{2012jet_plot}. The inner uncertainty band stands for the DSSV global QCD fit within a $90\%$ confidence level limit on the running integral of $\Delta g$ including the projected uncertainties up to the 2015 running period. 

\section{Summary and Outlook}
The latest results of the STAR jet measurements improve our understanding of both the unpolarized and polarized gluon distribution functions. The new inclusive jet cross section is expected to reduce the uncertainty on the gluon distribution function in high x region ($x>0.1$). A non-zero gluon spin contribution in the range of $0.05<x<1$ is indicated from the 2009 inclusive jet double spin asymmetry $A_{LL}$ result. The total gluon spin contribution could be $90 \%$ confidence level possibility of the same order as the quark polarization contribution to the proton spin. 

The uncertainties of the polarized gluon PDF at low $x$ remain large compared to the currently probed region. Longer 500 GeV longitudinally polarized proton-proton operations at RHIC and a possible forward upgrade will provide opportunities to measure di-jet production at STAR in the forward pseudorapidity with high precision data. Such observables are under study, and future analysis will impact the determination of the gluon helicity dependent PDFs and provide new constraints for the proton spin at low $x$ region.

\section{References}
%\bibliographystyle{iopart-num}
%\bibliography{xuanli}

\end{document}